\documentclass[twocolumn,showpacs,nofootinbib,amsmath,amssymb,aps,prd,superscriptaddress]{revtex4}
\usepackage{amsmath}
\usepackage{amsfonts,amssymb,mathrsfs}
\usepackage{hyperref}

\def\be{\begin{equation}}
\def\ee{\end{equation}}

\def\f{\frac}
\def\tf{\tfrac}

\def\pl{{\rm Pl}}
\def\lp{\ell_\pl}

\def\d{\dot}

\def\t{\tilde}

\def\wh{\widehat}

\def\bra{\langle}
\def\ket{\rangle}
\def\dd{{\rm d}}

\def\del{\partial}

\def\ga{\gamma}
\def\la{\lambda}
\def\om{\omega}

\def\De{\Delta}

\def\oq{\mathring{q}}

\def\mH{\mathcal{H}}
\def\mC{\mathcal{C}}
\def\mO{\mathcal{O}}
\def\mV{\mathcal{V}}

\usepackage{color}

\begin{document}

\pagestyle{plain}

\title{Why Are the Effective Equations of Loop Quantum Cosmology so Accurate?}

\author{Carlo Rovelli} \email{rovelli@cpt.univ-mrs.fr}
\affiliation{Aix Marseille Universit\'e, CNRS,
CPT, UMR 7332, 13288 Marseille, France}
\affiliation{Universit\'e de Toulon, CNRS, CPT,
UMR 7332, 83957 La Garde, France}

\author{Edward Wilson-Ewing} \email{wilson-ewing@phys.lsu.edu}
\affiliation{Department of Physics and Astronomy,
Louisiana State University, Baton Rouge, 70803, USA}

\begin{abstract}

We point out that the relative Heisenberg uncertainty relations
vanish for non-compact spaces in homogeneous loop quantum cosmology.
As a consequence, for sharply peaked states quantum fluctuations in
the scale factor never become important, even near the bounce
point.  This shows why quantum back-reaction effects remain
negligible and explains the surprising accuracy of the effective
equations in describing the dynamics of sharply peaked wave
packets.  This also underlines the fact that minisuperspace
models ---where it is global variables that are quantized---
do not capture the local quantum fluctuations of the geometry.

\end{abstract}


\maketitle

\section{Introduction}
\label{s.intro}

The loop quantum cosmology (LQC) effective equations provide
quantum-gravity corrections to the classical solutions of the
Friedmann cosmologies \cite{Taveras:2008ke, psvt},
but they are expected to break down when quantum gravity
effects become stronger.  This expectation arises since quantum
fluctuations of the geometry are expected to become important as
the space-time curvature nears the Planck scale, and when quantum
fluctuations become large, quantum back-reaction could become
important \cite{Bojowald:2012xy}.  However, despite this expectation,
numerical evidence shows that the effective equations (even when
neglecting all quantum back-reaction effects) provide an excellent 
approximation to the full dynamics of sharply peaked states,
including at the bounce point where quantum gravity effects are
strongest \cite{Ashtekar:2006wn, Corichi:2011rt}.  In this sense,
the effective equations are surprisingly accurate in the Planck
regime.  Why is quantum back-reaction negligible in the effective
dynamics?

A closely related question concerns the magnitude of quantum
fluctuations of the geometry at the bounce. Naively, one could
expect classical space-time to become ill-defined at the bounce,
because of the dominance of quantum gravity effects that may cause
large quantum fluctuations.  But this is not necessarily the case:
LQC predicts bouncing geometries where quantum fluctuations of the
scale factor at the bounce can be arbitrarily small. How is this
possible?

In this paper we offer an explanation of these facts, which
we believe may shed some light on the foundations of LQC
(and minisuperspace quantum cosmology in general). For the
sake of simplicity we focus on the case of a flat
Friedmann-Lema\^itre-Robertson-Walker (FLRW) universe
with a perfect fluid of constant equation of state.
It is easy to generalize the discussion to other matter
fields and also to other homogeneous cosmologies.  We
consider states which, far away from the Planck regime,
admit a clear semi-classical interpretation, i.e., states
that are sharply peaked in both the configuration and
momentum variables.

\section{Hamiltonian Formulation of Cosmology}
\label{s.ham}

In loop quantum cosmology, the fundamental variables are taken to be 
the Ashtekar-Barbero connection $A_a^i$ and the densitized triad $E^a_i$, 
which in the flat FLRW space-time can be parametrized as
\cite{Ashtekar:2003hd}
\be \label{vars1}
A_a^i = \t{c} \, (\dd x^i)_a, \qquad
E^a_i = \t{p} \, \sqrt{\oq} \left(\f{\del}{\del x^i}\right)^a,
\ee
where $\t{p} = a(t)^2$, with $a(t)$ being the usual scale factor
in the FLRW model (and ignoring a possible sign factor due to the
orientation of the triads), and $x^i$ are Cartesian coordinates
on the spatial manifold defining a fiducial spatial metric 
\be
\dd \mathring{s}^2 = \left( \dd x^1 \right)^2
+ \left( \dd x^2 \right)^2 + \left( \dd x^3 \right)^2,
\ee
with determinant $\oq=1$.

It is possible to rescale the fiducial coordinates by a factor $\alpha$,
$x^i \to \alpha x^i$ in which case $\oq \to \alpha^6 \oq$ and
\be \label{scale2}
\t{c} \to \alpha^{-1} \t{c}, \qquad
\t{p} \to \alpha^{-2} \t{p},
\ee
so that $A_a^i$ and $E^a_i$ remain invariant.

Fix a space-time region $\mV$, called the fiducial cell, with
fiducial-metric volume $V_o = \int_\mV \sqrt{\oq}$.
Inserting the form \eqref{vars1} for the variables $A_a^i$ and $E^a_i$,
the Holst action \cite{Holst:1995pc} for this region is given by the
constraints plus the ``symplectic" term
\be \label{symplectic}
\f{1}{8 \pi G \ga} \int_\mV \d{A_a^i} E^a_i \, \dd^3 x
= \f{3 V_o}{8 \pi G \ga} \, \d{\t{c}} \, \t{p},
\ee
where $\ga$ is the Immirzi parameter.  Notice that if space is
non-compact, for the theory to be well-defined, it is necessary
to introduce the fiducial cell $\mV$ in order to avoid the
divergence of this integral (and also the integrals that appear
in the constraint functions).  In this sense, $\mV$ acts as
an infrared regulator.

It follows that the fundamental Poisson bracket in the gravitational
sector is
\be \label{poisson}
\{ \t{c}, \t{p} \}  = \f{8 \pi G \ga}{3 V_o}.
\ee
This Poisson bracket is invariant under the rescaling \eqref{scale2},  
but it \emph{does depend on the choice of the fiducial cell} $\mV$,
as emphasized in \cite{Corichi:2011sd}.  This dependence does not
affect the classical dynamics, but it does play a very important
(though subtle) role in the quantum theory.

The dynamics is given by the Hamiltonian constraint $\mC_H$.  Taking
the lapse $N=1$, and in the presence of a massless scalar field
$\t\phi$ whose Poisson bracket with its conjugate momentum is
$\{\t\phi, \t\pi_\phi\} = V_o^{-1}$, $\mC_H$ is \cite{Ashtekar:2006wn}
\be \label{ham-con}
\mC_H = \int_\mV \, N \mH =
- \f{3 V_o \, \sqrt{\t{p}} \, \t{c}^2}{8 \pi G \ga^2}
+ V_o \, \f{\t\pi_\phi^2}{2 \t{p}^{3/2}} \approx 0.
\ee
From $\d\mO = \{\mO, \mC_H\}$, it follows that the classical
dynamics is independent from the fiducial cell: $V_o$
appears linearly in the Hamiltonian constraint and is cancelled
by the $1/V_o$ factor in the Poisson bracket \eqref{poisson}.

Since the classical dynamics is independent of the choice of the
fiducial cell, there is no need to remove the ``infrared regulator"
provided by the choice of the cell. In other words, the removal
of the regulator in the classical theory is trivial since the
dynamics is not affected by the choice of $\mV$.

\section{Quantum Cosmology}
\label{s.qc}

In the quantum theory, the Poisson brackets are replaced by commutators
and the variables become operators.  In order to gain some intuition, we
shall start by considering Wheeler-DeWitt (WDW) quantum cosmology as an
example.  In WDW theory, the basic commutator is given by
\be \label{comm}
[ \, \wh{\t{c}}, \wh{\t{p}} \, ]  = i \f{8 \pi G \hbar \ga}{3 V_o},
\ee
and at first sight, it seems as though the quantum theory could also
be independent from the choice of $\mV$ since in the Heisenberg 
picture
\be
\f{\dd \wh \mO}{\dd t} = \left[ \wh \mO, \wh{ \mC_H } \right],
\ee
and again the numerical factors of $V_o$ cancel out.

But the quantum theory does more than just giving the dynamics: it
also determines the quantum fluctuations of the classical variables. 
The commutation relation \eqref{comm} implies the uncertainty relation
\be
\De \t{c} \cdot \De \t{p} \ge \f{4 \pi G \hbar \ga}{3 V_o}.
\label{heisenberg}
\ee
This time nothing cancels the $V_o$ factor.  Therefore the choice 
of the quantisation region, i.e., of the fiducial cell, affects the
quantum theory.  (Another way to see that the quantum theory is not
invariant under a rescaling of the fiducial cell is by explicitly
determining how the expectation values of a given state change for
different choices of $V_o$ \cite{Corichi:2011sd}.)

In fact, by taking $V_o$ arbitrarily large, the right hand side of the
Heisenberg uncertainty relations \eqref{heisenberg} vanishes. Clearly,
the quantum fluctuations can then become arbitrarily small.  This does
not mean that all states must have small quantum fluctuations, but it
means that it is possible to construct states with arbitrarily small 
uncertainties in $\t{c}$ as well $\t{p}$.

The minimum possible amplitude of quantum fluctuations is a physical
quantity. If we require it to be independent of the fiducial cell,
because we want to view $\mV$ just as an infrared regulator, we can
take the limit of $V_o \to \infty$. This is routinely done in quantum
cosmology.  This gives
\be
\lim_{V_o \to \infty} \De \t{c} \, \De \t{p} \ge 0,
\ee
showing that there is no minimal non-zero amplitude for quantum
fluctuations in quantum cosmology, at least for non-compact
homogeneous spaces.  Thus, it is possible to build states whose
quantum fluctuations are always arbitrarily small.

It is clear that since the classical symplectic structure in LQC is
the same as for WDW quantum cosmology, similar arguments will apply
for LQC as well, and this explains why the effective equations of LQC
are so accurate.  Even more, it is possible to go further and show
that for large $V_o$, states (in the Schr\"odinger picture) whose
quantum fluctuations are small at one time continue to have small
quantum fluctuations at all times, including at the bounce point.
This result implies (for large $V_o$) that since the quantum
fluctuations remain small, the effective Hamiltonian is expected
to give an excellent approximation to the full quantum dynamics at
all times, including the bounce point.

In order to show this in LQC, it is convenient to change variables to
\cite{Ashtekar:2007em}
\be \label{vars}
\t\beta = \f{\t{c}}{\sqrt{\t{p}}}, \qquad \t{V} = \t{p}^{3/2},
\ee
whose Poisson bracket, in the classical theory as well as
the effective theory, is
\be
\{ \t\beta, \t{V} \} = \f{4 \pi G \ga}{V_o}.
\ee

In LQC, all expressions that contain the connection must be expressed in
terms of holonomies.  Because of this, there is no operator corresponding
to $\t\beta$; rather the operators of interest are complex exponentials (or
trigonometric functions) of $\ell \t\beta$, which correspond to holonomies
of the Ashtekar-Barbero connection along edges of \emph{physical} length
equal to $\ell$.  In particular, in the Hamiltonian constraint operator
the field strength operator is expressed in terms of holonomies around a
minimal area loop with $\ell = \la$, where $\la^2 \sim \lp^2$ is the area gap,
the smallest non-zero eigenvalue of the area operator in loop quantum gravity
\cite{Ashtekar:2006wn}.  As shall be seen below, this input appears in the
resulting LQC effective theory.

Now we shall briefly recall the main ingredients of LQC that will be necessary
here, for more details see \cite{Ashtekar:2006wn, Ashtekar:2007em}.  A convenient
basis for the LQC of the flat FLRW space-time are eigenstates of the volume operator,
\be \label{vol-op}
\wh{\t{V}} \, |\t{V}\ket = \t{V} \, |\t{V}\ket, \quad
{\rm with} \quad \bra \t{V}_1 | \t{V}_2 \ket = \delta_{\t{V}_1,\t{V}_2}.
\ee
The other basic operators in the gravitational sector of LQC are the holonomy (also
called shift) and inverse volume operators, defined as
\begin{align}
\label{shift-op}
\wh{e^{-i \ell \t\beta}} \, | \t{V} \ket &=
| \t{V} + 4 \pi G \hbar \ga \ell / V_o \ket, \\
\label{inv-v}
\wh{ \t{V}^{-1} } | \t{V} \ket &=
\begin{cases}
0 & \text{if  } \t{V} = 0,\\
\t{V}^{-1} \: | \t{V} \ket & \text{otherwise.}\\
\end{cases}
\end{align}
Note that there exists a large number of ambiguities in the definition of inverse
volume operators (see e.g.\ Sec.\ IV in \cite{Singh:2013ava} for a more detailed
discussion on this point), here we choose \eqref{inv-v} both for its simplicity,
and because it is the only known inverse volume operator in LQC that does not
depend on the choice of the fiducial cell.

From \eqref{vol-op} and \eqref{shift-op}, it follows that the basic commutator
in LQC is
\be \label{lqc-comm}
\left[ \wh{\t{V}}, \wh{e^{-i \ell \t\beta}} \right]
= \f{4 \pi G \hbar \ga \ell}{V_o} \, \wh{e^{-i \ell \t\beta}}.
\ee
At this point, it is possible to define the Hamiltonian constraint operator
in LQC and then study the resulting quantum dynamics, as in
\cite{Ashtekar:2006wn, Ashtekar:2007em}.

Instead, here we shall use the effective equations and, assuming a state
that is initially sharply peaked in both its configuration and momentum variables,
determine when quantum fluctuations become comparable to expectation values.
It is at this point that quantum backreaction will become important and the
effective theory can no longer be trusted.  Thus, the effective theory itself
will tell where it breaks down.

In the $(\t{V}, \t{\beta})$ variables, the LQC effective theory is determined
by the effective Hamiltonian constraint \cite{Taveras:2008ke}
\be \label{ch-eff}
\mC_H = - \f{3 V_o \, \t{V}}{8 \pi G \ga^2 \la^2}
\sin^2 \left( \la \t\beta \right)
+ V_o \, \t{V} \, \rho \approx 0,
\ee
for a generic perfect fluid.  The continuity equation is unchanged in
the effective theory of LQC (recall $\t{V} = a^3$)
\be
\f{\dd \rho}{\dd t} + \f{1}{\t{V}} \f{\dd {\t{V}}}{\dd t} \left( \rho + P \right) = 0.
\ee
Assuming a constant equation of state $P = \om \rho$ with $-1 \le \om \le 1$,
\be
\rho = \f{\t\rho_o}{\t{V}^n}, \qquad 0 \le n \le 2,
\ee
with $n = 1 + \om$.  Note that since $\rho$ is independent of $V_o$,
$\t\rho_o$ is as well.

The commutator \eqref{lqc-comm} implies the uncertainty relation
\be \label{lqc-un}
\Delta \t{V} \cdot \Delta \left( \f{\sin \la \t\beta}{\la} \right)
\ge \f{2 \pi G \hbar \ga}{V_o} \cdot \left| \bra \cos \la \t\beta \ket \right|,
\ee
and from this relation it is possible to calculate when quantum fluctuations
become important.  The constraint \eqref{ch-eff} determines the value of $\t{V}$
when the bounce occurs in the effective theory and so it is possible to check
whether the quantum fluctuations become important before the bounce occurs
or not.

Hamilton's equation for $\t{V}$ from \eqref{ch-eff} is
\be
\f{\dd{\t{V}}}{\dd t} = \f{3 \t{V}}{\ga \la} \sin \left( \la \t\beta \right)
\cos \left( \la \t\beta \right);
\ee
this equation of motion can be solved by squaring it and then using the
constraint equation $\mC_H = 0$, giving
\be
\t{V}(t) = \left( 6 \pi G \t\rho_o n^2 \left( t - t_o \right)^2
+ \f{\t\rho_o}{\rho_c} \right)^{1/n},
\ee
where the critical energy density is $\rho_c = 3/(8 \pi G \ga^2 \la^2)$.
The resulting $\t{V}_{\rm bounce}$ is independent of $V_o$,
\be
\t{V}_{\rm bounce} = \left( \f{\t\rho_o}{\rho_c} \right)^{1/n}.
\ee

Now let us determine the volume where initially small quantum fluctuations
become important.  $\mC_H = 0$ implies
\be \label{sin-lb}
\f{\sin \la \t\beta}{\la} = \f{\sqrt{\t\rho_o}}{\la \sqrt{\rho_c}}
\cdot \f{1}{\t{V}^{n/2}}.
\ee
Then, as $\Delta[f(x)] = \Delta(x) \cdot | \partial_x f(x)|$ for small
$\Delta x$, we find that%
\footnote{Here we are neglecting the fluctuations
in the matter field.  This is because the matter
field is typically used as a relational clock in
LQC and so acts as a parameter with respect to
which other observables are measured.  Note that
the qualititative result of this calculation
is not affected if we include matter fluctuations
as well.}
\be \label{db-dv}
\Delta \left( \f{\sin \la \t\beta}{\la} \right) =
\f{n \sqrt{\t\rho_o}}{2 \la \sqrt{\rho_c}} \f{\Delta \t{V}}{\t{V}^{n/2+1}}.
\ee
Note that here we are only considering states that are initially
sharply-peaked, by which we mean that initially the relative uncertainties
$\Delta \t{V}/\t{V}$ and $\Delta (\sin \la \t\beta) / \sin \la \t\beta$
are small compared to 1, and that the higher-order moments are smaller
still.

Assuming that the uncertainty relation \eqref{lqc-un} is nearly
saturated and using \eqref{db-dv}, we find
\be
\f{\left( \Delta \t{V} \right)^2}{\t{V}^{n/2+1}}
\sim \f{4 \pi G \hbar \ga \la}{n V_o} \sqrt \f{\rho_c}{\t\rho_o}
\, \left| \cos \la \t\beta \right|.
\ee

Quantum fluctuations become important when $\t{V} \sim \Delta \t{V}$,
which occurs for
\be
\t{V}_{\rm qf} \sim \left( \f{4 \pi G \hbar \ga \la}{n V_o}
\sqrt \f{\rho_c}{\t\rho_o} \, \left| \cos \la \t\beta \right|
\right)^{\tf{2}{2-n}},
\ee
and we immediately see that, for large $V_o$,
\be \label{inequality}
\t{V}_{\rm bounce} \gg \t{V}_{\rm qf}.
\ee
An analogous calculation can be performed to determine
$\sin (\la \t\beta)_{\rm bounce}$ and
$\sin (\la \t\beta)_{\rm qf}$, this gives
\be
\sin (\la \t\beta)_{\rm bounce} = 1,
\ee
and
\be
\sin (\la \t\beta)_{\rm qf} \sim V_o^{\tf{n}{2-n}};
\ee
where we have only written the $V_o$ dependence for
$\sin (\la \t\beta)_{\rm qf}$.  Clearly, as $2-n \ge 0$, it is
impossible for $\sin (\la \t\beta)_{\rm qf}$ to be reached if
$V_o$ is taken to be sufficiently large.

Thus, it is clear that, for large $V_o$, states that are initially
sharply peaked will remain sharply peaked throughout their evolution
as they bounce before quantum fluctuations have a chance to become
important, and the effective Hamiltonian will provide an excellent
approximation to the full quantum dynamics at all times.

Of course, it is important to keep in mind that for states that are
not sharply peaked, the effective equations are not a good approximation
as can be seen explicitly in \cite{Diener:2013uka}.  In addition, for
small $V_o$, we see that $\t{V}_{\rm qf}$ may be larger than
$\t{V}_{\rm bounce}$ in which case the effective dynamics generated
by the effective Hamiltonian \eqref{ch-eff} cannot be trusted for
$\t{V} \lesssim \t{V}_{\rm qf}$.  Thus, the effective equations
may fail for states that are not sharply peaked, or where $V_o$
is small.

Very similar calculations yield the same results for compact models:
the effective equations provide an excellent approximation to the
full quantum dynamics for sharply peaked states so long as $V_o$
is sufficiently large.  [One difference in the calculation is that
in a compact space, it is not necessary to introduce a fiducial
cell as the integrals \eqref{symplectic} and \eqref{ham-con} are
bounded and so do not diverge.  Then, $V_o$ corresponds to the
volume of the entire space with respect to $\oq_{ab}$ and therefore
$V_o$ is no longer a free parameter but rather is fixed.  Another
point is that other inverse triad operators than \eqref{inv-v} may
be chosen in compact spaces; while this would complicate the
calculations, the qualitative results should remain unchanged.
Other than these points, the calculation for the compact space
is essentially identical to the one given here for a non-compact
space.]  Thus, we expect sharply peaked states in compact
spaces to also remain sharply peaked at all times, including at
the bounce point, so long as $V_o$ is sufficiently large.  On the
other hand, for small $V_o$, quantum fluctuations are expected to
become important before the bounce occurs, in which case the
effective equations may no longer provide a good approximation
to the full quantum dynamics.

Now it is clear why the effective equations are so accurate, even
in the deep Planck regime: in order to build a semi-classical state,
fluctuations must be small in the classical limit.  Since there is
no lower bound on the amplitude of quantum fluctuations in the
theory, it is possible to choose states where the quantum fluctuations
are arbitrarily small, and so never become important, even at the
bounce point.

It has been pointed out that, in general, quantum back-reaction from
higher order moments must be included in the effective equations
\cite{Bojowald:2012xy}, but quantum back-reaction can safely be ignored
so long as quantum fluctuations are negligible.  As we have seen
here, there exists a large family of solutions (i.e., solutions with
initially small fluctuations and sufficiently large $V_o$) where
ignoring quantum back-reaction is a valid approximation.

This is also why the fluctuations in the scale factor and other
large-scale observables can be negligible: for non-compact spaces in
LQC (and also WDW and other quantum cosmology minisuperspace
models) it is always possible, in the limit $V_o \to \infty$, for
the fluctuations to be arbitrarily small.

These considerations answer the two questions we asked in
the introduction.  Still, the answer is a little disconcerting.
What is the sense of a quantum theory where fluctuations can
always be reduced to zero?  Isn't this in contradiction with
standard quantum theory ideas, and the irreducibility of the
Heisenberg relations?

\section{Where Are the Fluctuations?}
\label{s.where}

A source of confusion in searching for a description of the quantum 
fluctuations in cosmology is due to a common misinterpretation of 
the theory.  When we restrict the space of the solutions of the classical
theory to the homogenous fields \eqref{vars}, and we consider only their
dynamics, we are effectively disregarding the quantum theory of all of
the higher modes of the field. The commutation relations \eqref{comm}
show that the larger modes behave effectively as averages of local
variables.  

An analogy of this situation is provided by the following
example.  Consider a material formed by $N$ atoms, each having unit
mass. Let the position $x_n$ and momentum $p_n=\dot x_n$ of each atom 
satisfy standard commutation relations $[x_n,\dot x_n]=i\hbar$.
The center of mass of the system
\be
X = \frac1N \sum_n x_n
\ee
and its velocity $\dot X$ satisfy the commutation relation
\be \label{comm-cm}
[X, \dot X] = \frac{i\hbar}N ,
\ee
which goes to zero as $N$ goes to infinity.  Thus the centre of mass 
effectively behaves classically when there are many particles.  
On physical grounds, this obviously does not imply that the
quantum fluctuations of the individual atoms go to zero in the large
$N$ limit. It only implies that the centre of mass is blind to these.
A more general discussion on the classical limit of macroscopic
(large $N$) observables can be found in \cite{Poulin:2004}.

A moment of reflection shows that the mathematics of quantum 
cosmology is analogous. The homogenous variables are the lowest 
modes in a Fourier expansion and are therefore like averages of the
local variables, namely the fields at a point.   As before, the fact that 
the commutation relations \eqref{comm} vanish in the large $V_o$ limit
does not imply at all that the quantum fluctuation of the fields at a point
go to zero. It only shows that $\t\beta$ and $\t{V}$ are blind to these. 
In other words, homogeneous quantum cosmology is blind to the local 
quantum fluctuations of the gravitational field. 

The theory with a given fiducial cell $\cal V$, in other words, describes
only the quantum effects on the modes of the size of $\cal V$, and
not the smaller modes.   Different choices of  $\cal V$ do not lead to
a mathematical lack of definiteness: they correspond to analyzing different
modes of the theory.  For example, in order to study the dynamics of
the large-scale structure of the universe, it is appropriate to choose a
very large fiducial cell, in which case quantum fluctuations will be
negligible for sharply-peaked states.  On the other hand, we do expect
quantum fluctuations to necessarily be important for the physics at
trans-Planckian scales.  We will discuss this last point further in
Sec.\ \ref{s.disc}.

To know how the geometry fluctuates at short scales, we can simply take
$\cal V$ to be small rather than large.  It is this that provides a
description of local quantum fluctuations.  Let us sketch here what
this implies. 

To do this calculation, it is convenient to use the variables
\be
V = V_o \, \t{V}, \qquad \beta = \t\beta,
\ee
where $V$ corresponds to the physical volume of the fiducial cell
\cite{Ashtekar:2006wn, Ashtekar:2007em}.

As an aside, note that this rescaling is akin to using the centre of mass
momentum $P = N \d X$ in \eqref{comm-cm} rather than the velocity.  In this
case the basic commutator is $[X, P] = i\hbar$, which is independent of $N$.
Of course, using different variables does not change the underlying physics
in the many-body example (or in LQC) and the centre of mass variable effectively
behaves classically for large $N$, whether one uses the variable $\d X$ or $P$.

Similarly, the Poisson bracket in the new variables is given by
$\{V, \beta\} = 4 \pi G \ga$ and therefore the uncertainty relations in the
quantum theory are independent of $V_o$.  Nonetheless, we can easily show that
the same result holds.  The basic uncertainty relation in terms of the
new variables is
\be
\Delta V \cdot \Delta(\sin \la\beta) \ge 2 \pi G \hbar \ga \la \cos \la\beta.
\ee
In order to bound the relative uncertainties, we divide both sides by
$V \sin \la\beta$, and on the right-hand side use the relation
$\sin \la\beta = \sqrt{\rho / \rho_c}$ given in \eqref{sin-lb},
\be \label{comm-vb}
\f{\Delta V}{V} \cdot \f{\Delta(\sin \la\beta)}{\sin \la\beta}
\ge 2 \pi G \hbar \ga \la \,
\f{\sqrt{\rho_c} \cos \la\beta}{\sqrt{\rho_o} \, V^{(2-n)/2}},
\ee
where $\rho_o = V_o^n \t{\rho_o}$.  It is easy to check that while the
left-hand side of \eqref{comm-vb} is independent of $\mV$, the right-hand
side goes as $1/V_o$.

Furthermore, bounding the $\cos\la\beta$ term by 1,
the strongest lower bound on the relative uncertainties occurs for the smallest
value of $V$ that is reached, which is the volume of the fiducial cell at the
bounce $V_{\rm bounce}$. At the bounce point, the effective theory predicts
that $\rho = \rho_c$, which implies that $\rho_o = \rho_c \cdot V_{\rm bounce}^n$.
Therefore the strongest lower bound on the relative uncertainties is
\be
\f{\Delta V}{V} \cdot \f{\Delta(\sin \la\beta)}{\sin \la\beta}
\ge \f{2 \pi G \hbar \ga \la}{V_{\rm bounce}}.
\ee
This shows that so long as the physical volume of the fiducial cell at the
bounce point is significantly larger than the Planck volume, the relative
uncertainties will never become large for an initially sharply-peaked state.
Note however that if one chooses a fiducial cell so that $V_{\rm bounce} \ge
2 \pi \ga \lp^2 \la$ (recall $\la \sim \lp$), then quantum fluctuations will
become important when $V \sim 2 \pi \ga \lp^2 \la$ at which point the effective
equations will no longer be reliable.

Furthermore, it is also clear how this result generalizes to compact
space-times: so long as the physical volume of the spatial slice remains
much larger than the Planck volume, the relative uncertainties will not become
important and the effective equations will be accurate.

\section{Conclusions}
\label{s.disc}

We have concentrated on the flat FLRW model, but the
results presented here can easily be generalized to other homogeneous
cosmologies, whether the spatial manifold is compact or non-compact.
This indicates that, for sharply peaked states, quantum back-reaction
effects are negligible and the effective equations derived for the
loop quantum cosmology of the Bianchi space-times and the
Kantowski-Sachs space-times should be an excellent approximation to
the quantum dynamics, so long as all length scales in the space-time
remain much larger than the Planck length at all times.

These results can also be applied to cosmological perturbation theory.
One way to study perturbations (up to some minimal wavelength) is by
dividing the spatial manifold into a lattice of cells, where each cell
is taken to be homogeneous.  As the gravitational and matter fields
vary from cell to cell, the fields are inhomogeneous at scales larger
than the size of the cells.  If the fields from one cell to another
vary slightly around some mean value (which is taken to be the homogeneous
background), then this setting can be used to study cosmological
perturbations \cite{WilsonEwing:2012bx}.  Since each cell is assumed
to be homogeneous, the arguments given here indicate that effective
equations can be trusted (again assuming sharply peaked states) so
long that the physical volume of each cell remains significantly larger
than the Planck volume.  As the minimal wavelength captured by
this lattice is given by (twice) the cube root of the volume of each cell,
this corresponds to modes whose wavelengths are much larger than the
Planck length at all times.  Therefore, the effective equations
will be valid for such modes; however, they cannot be expected to be
a good approximation for trans-Planckian modes where quantum fluctuations
will necessarily be important.

This example further emphasizes that the choice of the fiducial cell
must be made by taking into account the physics of interest.  If one wishes
to study the dynamics of the scale factor, the mean space-time curvature,
the mean energy density and other large-scale observables that are typically
of interest in minisuperspace models, it is appropriate to choose a very large
fiducial cell and, for non-compact spaces, even take the limit of $V_o \to \infty$.
On the other hand, this analysis has also shown that quantum fluctuations
will necessarily be important for trans-Planckian physics, and we do not
expect effective equations to be reliable in that setting.

In short, \emph{local} quantum fluctuations of the metric are not
captured by minisuperspace models.  The scale factor is a global quantity
where the local quantum fluctuations are largely averaged out.  This is why,
for states that are initially sharply peaked, and so long as the volume of
the region under consideration remains much larger than $\lp^3$, relative
quantum fluctuations can be arbitrarily small in minisuperspace models such
as LQC.

\acknowledgments

We thank Parampreet Singh for
helpful discussions.
This work was supported in part by
a grant from the John Templeton Foundation.
The opinions expressed in this publication are those of the
authors and do not necessarily reflect the views of the John
Templeton Foundation.

%

\begin{thebibliography}{10}
\raggedright


\bibitem{Taveras:2008ke}
V.~Taveras, ``{Corrections to the Friedmann Equations from LQG for a Universe
  with a Free Scalar Field},'' Phys.\ Rev.\ {\bf D78} (2008) 064072,
\href{http://arXiv.org/abs/0807.3325}{{\tt arXiv:0807.3325}}.

\bibitem{psvt}
P.~Singh and V.~Taveras, ``{A note on the effective equations in LQC},'' to be
  published.

\bibitem{Bojowald:2012xy}
M.~Bojowald, ``{Quantum Cosmology: Effective Theory},'' Class.\ Quant.\ Grav.\
  {\bf 29} (2012) 213001,
\href{http://arXiv.org/abs/1209.3403}{{\tt arXiv:1209.3403}}.

\bibitem{Ashtekar:2006wn}
A.~Ashtekar, T.~Paw{\l}owski, and P.~Singh, ``{Quantum Nature of the Big Bang:
  Improved dynamics},'' Phys.\ Rev.\ {\bf D74} (2006) 084003,
\href{http://arXiv.org/abs/gr-qc/0607039}{{\tt arXiv:gr-qc/0607039}}.

\bibitem{Corichi:2011rt}
A.~Corichi and E.~Montoya, ``{Coherent semiclassical states for loop quantum
  cosmology},'' Phys.\ Rev.\ {\bf D84} (2011) 044021,
\href{http://arXiv.org/abs/1105.5081}{{\tt arXiv:1105.5081}}.

\bibitem{Ashtekar:2003hd}
A.~Ashtekar, M.~Bojowald, and J.~Lewandowski, ``{Mathematical structure of loop
  quantum cosmology},'' Adv.\ Theor.\ Math.\ Phys.\ {\bf 7} (2003) 233--268,
\href{http://arXiv.org/abs/gr-qc/0304074}{{\tt arXiv:gr-qc/0304074}}.

\bibitem{Holst:1995pc}
S.~Holst, ``{Barbero's Hamiltonian derived from a generalized Hilbert-Palatini
  action},'' Phys.\ Rev.\ {\bf D53} (1996) 5966--5969,
\href{http://arXiv.org/abs/gr-qc/9511026}{{\tt arXiv:gr-qc/9511026}}.

\bibitem{Corichi:2011sd}
A.~Corichi and E.~Montoya, ``{On the Semiclassical Limit of Loop Quantum
  Cosmology},'' Int.\ J.\ Mod.\ Phys.\ {\bf D21} (2012) 1250076,
\href{http://arXiv.org/abs/1105.2804}{{\tt arXiv:1105.2804}}.

\bibitem{Ashtekar:2007em}
A.~Ashtekar, A.~Corichi, and P.~Singh, ``{Robustness of key features of loop
  quantum cosmology},'' Phys.\ Rev.\ {\bf D77} (2008) 024046,
\href{http://arXiv.org/abs/0710.3565}{{\tt arXiv:0710.3565}}.

\bibitem{Singh:2013ava}
P.~Singh and E.~Wilson-Ewing, ``{Quantization ambiguities and bounds on
  geometric scalars in anisotropic loop quantum cosmology},'' Class.\ Quant.\ Grav.\
  {\bf 31} (2014) 035010,
\href{http://arXiv.org/abs/1310.6728}{{\tt arXiv:1310.6728}}.

\bibitem{Diener:2013uka}
P.~Diener, B.~Gupt, and P.~Singh, ``{Chimera: A hybrid approach to numerical
  loop quantum cosmology},'' Class.\ Quant.\ Grav.\ {\bf 31} (2014) 025013,
\href{http://arXiv.org/abs/1310.4795}{{\tt arXiv:1310.4795}}.

\bibitem{Poulin:2004}
D.~Poulin, ``{Macroscopic Observables},'' Phys.\ Rev.\ {\bf A71} (2005) 022102,
  \href{http://arXiv.org/abs/quant-ph/0403212}{{\tt arXiv:quant-ph/0403212}}.

\bibitem{WilsonEwing:2012bx}
E.~Wilson-Ewing, ``{Lattice loop quantum cosmology: scalar perturbations},''
  Class.\ Quant.\ Grav.\ {\bf 29} (2012) 215013,
\href{http://arXiv.org/abs/1205.3370}{{\tt arXiv:1205.3370}}.

\end{thebibliography}
%

\end{document}